\documentclass{PoS}
\usepackage{epsfig}

\newif\ifpreprint
\preprinttrue
\newcommand{\BlackHat}{{\sc BlackHat}}

\newcommand{\SHERPA}{{\sc SHERPA}}

\newcommand{\COMIX}{{\sc COMIX}}

\newcommand{\ntuples}{{$n$-tuples}}

\def\eqn#1{eq.~(\ref{#1})}

\def\eqns#1#2{eqs.~(\ref{#1}) and~(\ref{#2})}

\def\nn{\nonumber}

\def\WZjjj{$W,Z\,\!+\,3$}

\def\Wjnp1{$W\,\!+\,(n\!+\!1)$}

\def\Zjn{$Z\,\!+\,n$}

\def\WZjjj{$W,Z/\gamma^*\,\!+\,3$}

\def\YYjj{$\gamma\gamma\,\!+\,2$}
\def\YYj{$\gamma\gamma\,\!+\,1$}
\def\YY{$\gamma\gamma\,\!+\,0$}
\def\gjn{$\gamma\,\!+\,n$}
\def\jet{{\rm jet}}
\def\pT{p_{\rm T}}

\def\root{{\sc root}}

\def\HTpartonic{{\hat H}_{\rm T}}

\def\pspacer{\hfil\rule{1cm}{0cm}\hfil}
\title{%
\ifpreprint
\vskip -1 cm 
\hbox{
\small\rm UCLA/13/TEP/110\pspacer
SLAC--PUB--15847\pspacer
SB/F/430-13 \pspacer
IPhT--T13/269\pspacer
IPPP/13/96
}\fi
Next-to-leading order diphoton+2-jet production at the LHC}

\ShortTitle{Next-to-leading order diphoton+2-jet production at the LHC}

\author{Z.~Bern\\
        Department of Physics and Astronomy, UCLA, Los Angeles, CA 90095-1547, USA\\
        E-mail: \email{bern@physics.ucla.edu}}

\author{L.~J.~Dixon\\
        SLAC National Accelerator Laboratory, Stanford University, Stanford, CA 94309, USA \\
        E-mail: \email{lance@slac.stanford.edu}}

\author{F.~Febres Cordero\\
        Departamento de F\'{\i}sica, Universidad Sim\'on Bol\'{\i}var,  Caracas 1080A, Venezuela\\
        E-mail: \email{ffebres@usb.ve}}
\author{S.~H{\"{o}}che\\
        SLAC National Accelerator Laboratory, Stanford University, Stanford, CA 94309, USA \\
        E-mail: \email{shoeche@slac.stanford.edu}}

\author{H.~Ita\\
        Physikalisches Institut, Albert-Ludwigs-Universit\"at Freiburg,
       D--79104 Freiburg, Germany\\
        E-mail: \email{harald.ita@physik.uni-freiburg.de}}

\author{D.~A.~Kosower\\
        Institut de Physique Th\'eorique, CEA--Saclay,
          F--91191 Gif-sur-Yvette cedex, France\\
        E-mail: \email{david.kosower@cea.fr}}

\author{\speaker{N. A. Lo Presti}\\
        Institut de Physique Th\'eorique, CEA--Saclay,
          F--91191 Gif-sur-Yvette cedex, France\\
        E-mail: \email{nicola.lo-presti@cea.fr}}

\author{D.~Ma\^{\i}tre\\
        Institute for Particle Physics Phenomenology, University of Durham, Durham DH1 3LE, UK\\
        E-mail: \email{daniel.maitre@durham.ac.uk}}

\abstract{We present 
results from a recent calculation of
prompt photon-pair production in association
with two jets to next-to-leading order (NLO) at the LHC.
The virtual contribution is evaluated using the \BlackHat{} library, 
a numerical implementation of on-shell methods
for one-loop amplitudes, in conjunction with 
\SHERPA.
We study four sets of cuts: standard jet cuts, a set of
Higgs-related cuts suggested by ATLAS, and corresponding sets
which isolate the kinematic region where 
the process becomes the largest background
to Higgs production via vector-boson fusion.}

\FullConference{11th International Symposium on Radiative Corrections (Applications of Quantum Field Theory to Phenomenology) (RADCOR 2013),\\
		22-27 September 2013\\
		Lumley Castle Hotel, Durham, UK}

\begin{document}

\section{Introduction}

The QCD production of a photon pair in association with jets is an
important process at hadron colliders.  Photon pairs are one of the
key decay channels for detecting and measuring the Higgs-like
boson~\cite{AtlasHiggs,CMSHiggs} announced last year.  Accordingly, a
good understanding of prompt photon-pair production is important to
precision measurements of its properties and thence to exploring
deviations from Standard Model expectations.  In particular, when the
photon pair is produced in association with two hadronic jets, the
process is an important background to Higgs-like boson production via
vector-boson fusion (VBF).  Leading-order (LO) predictions in QCD
suffer from a strong dependence on the unphysical renormalization and
factorization scales. Next-to-leading order (NLO) predictions
generally reduce this dependence, and offer the first quantitatively
reliable predictions in perturbation theory.  Inclusive diphoton
production at hadron colliders was computed long ago at NLO
\cite{DIPHOX,TwogammaMC,MCFMPhoton} and even beyond \cite{Catani2011qz}.  The
production of a photon pair in association with a single jet has also
previously been calculated at
NLO~\cite{DelDuca2003uz,Gehrmann2013aga}.  Here we present
predictions for inclusive diphoton production in association with two
jets at NLO, which has also been studied recently by Gehrmann,
Greiner, and Heinrich~\cite{Gehrmann2013bga}.

\section{$\gamma\gamma+2$ jets at NLO}

The ingredients that enter a NLO calculation 
are the Born cross section, the virtual  (one-loop)
corrections, and the radiative (real-emission) corrections.
The latter are computed from tree-level matrix elements with an
additional parton in the final state compared to the Born process.
In order to implement in a numerical
setting the cancellation of the separate divergences
that arise in the virtual and real corrections, 
we use the Catani-Seymor dipole subtraction scheme
\cite{CS},
which introduces a fourth, subtraction, contribution
to the evaluation of the NLO cross section
 \begin{equation}
 \sigma_n^{{\rm NLO}}\,\,=\,\,
   \int_n \sigma_n^{{\rm born}}
   \,+\,
 \int_n \sigma_n^{{\rm virt}}
    \,+\, \int_n\Sigma_n^{{\rm subtr}}
   \,+\,
 \int_{n+1} \left(\sigma_{n+1}^{{\rm real}}
    \,-\, \sigma_{n+1}^{{\rm subtr}}\right)\,.
 \end{equation}
We use the Frixione photon isolation criterion~\cite{FrixioneCone},
which avoids the need for fragmentation contributions.

We use the SHERPA package~\cite{Sherpa} to manage the partonic subprocesses, 
to integrate over phase space, and to output \root{}~\cite{ROOT} \ntuples.
For the computation of the Born and real-emission matrix elements along with the Catani-Seymour dipole subtraction terms, we used the~\COMIX{} library~\cite{Comix}, which is included in
the \SHERPA{} framework.  It is based on a color-dressed form~\cite{CDBG} of the Berends--Giele
recursion relations~\cite{BG}.

We use the \BlackHat{} software library~\cite{BlackHatI,W3jPRL}
to evaluate the virtual contribution.  This library has previously been used
to evaluate the virtual contributions of a number of other LHC processes of interest,
such as  \WZjjj- \,and $4$-jet production~\cite{W3jDistributions,BlackHatZ3jet,W4j,Z4j},  four-jet
production~\cite{FourJets}, as well as in investigations
of high-$\pT$ $W$ polarization~\cite{Wpolarization}, and for a study of \gjn-jet to \Zjn-jet ratios~\cite{PhotonZ,PhotonZ3}.  

The techniques implemented numerically in the \BlackHat{} library are
known as on-shell methods, and are reviewed in
refs.~\cite{OnShellReviews} (for other numerical implementations of
on-shell methods see refs.~\cite{OtherOnShellPrograms,GOSAM,HELAC};
for other recent developments see refs.~\cite{OtherMethods}).
One-loop amplitudes in QCD with massless quarks may be expressed as a
sum over three different types of Feynman integrals (boxes, triangles,
and bubbles) with additional rational terms.  Since the integrals are
universal and well-tabulated, the aim of the calculation is to compute
their coefficients, along with the rational terms.  In an on-shell
approach, the integral coefficients may be computed using
four-dimensional generalized unitarity~\cite{UnitarityMethod, Zqqgg,
  NewUnitarity}, while the rational terms may be computed either by a
loop-level version~\cite{GenHel} of on-shell recursion~\cite{BCFW} or
using $D$-dimensional unitarity~\cite{DDimUnitarity}.  We use a
numerical version~\cite{BlackHatI} of Forde's method~\cite{Forde} for
the integral coefficients, and subtract box and triangle integrands in
a fashion similar to the Ossola--Papadopoulos--Pittau
procedure~\cite{OPP}, in order to improve the numerical stability of
the calculation.  To compute the rational terms, we use a numerical
implementation of Badger's massive-integrals method~\cite{Badger},
related to $D$-dimensional unitarity.

A \BlackHat{} numerical calculation of the virtual contribution
proceeds as follows: it calculates the fundamental building blocks
(the so-called primitive amplitudes~\cite{Primitive}); assembles them
into color-ordered amplitudes; and finally computes the interference
between tree and one-loop amplitudes, multiplied by the appropriate
color factors.  In the results reported here, we drop the
subleading-color virtual terms; we have checked that they amount to
under 2\% of the total cross section.  The five light quarks
($u,d,c,s,b$) are all treated as massless; we neglect contributions
from real or virtual top quarks.  In this calculation we include the
`pure-gluonic' contribution $\,gg\rightarrow\gamma\gamma gg\,$.  This
contribution is nominally of higher order, but the analogous
contributions have typically been included for photon-pair production
without jets or in association with one jet, because there is no
corresponding tree-level gluon-initiated process, and because the
large value of the gluon distribution can compensate for the
contribution's additional powers of the strong coupling $\alpha_s$.
With two associated jets, there {\it is\/} a tree-level process
($gg\rightarrow q\bar q\gamma\gamma$), and we would expect the
`pure-gluonic' contribution to be suppressed.  This is indeed what we
find; it contributes less than 3\% of the total cross section at NLO.

We have performed a number of checks on the virtual contributions for
photon-pair production in the \BlackHat{} library.  We have checked their
collinear factorization properties.  We have checked the matrix elements
at individual phase-space points against HELAC~\cite{HELAC} 
and MCFM~\cite{MCFM} values 
for \YY-jet production; against GoSam~\cite{GOSAM} for \YYj-jet
production; and against a previous analytic calculation~(relying on
\cite{qqggg}) for the 
specific subprocess $qg \rightarrow \gamma\gamma q $.
We compared selected helicity configurations at individual phase-space
points against GoSam for \YYjj-jet production.
We have also checked integrated results against MCFM
for \YY-jet production, and against Gehrmann, Greiner, and Heinrich's
results~\cite{Gehrmann2013aga} for \YYj-jet production.

\section{Kinematics and Observables}

In our study, we consider the inclusive process $p p \rightarrow\,\gamma\gamma\,+\,2\,$
 jets at an LHC center-of-mass energy of $\sqrt{s} = 8$ TeV, applying the following cuts:
\begin{eqnarray}
    && \pT^{\gamma_1} > 50 \hbox{ GeV} \,, \hskip 1.2 cm
\pT^{\gamma_2} > 25 \hbox{ GeV} \,, \hskip 1.2 cm 
|\eta^{\gamma}| < 2.5\,, \hskip 1.2 cm \nn\\
&& \pT^{\jet_1} > 40 \hbox{ GeV}\,, \hskip 1.2 cm 
   \pT^{\jet_2} > 25 \hbox{ GeV}\,, \hskip 1.2 cm 
  |\eta^{\jet}| < 4.5\,, \hskip 1.2 cm 
R_{\gamma,\jet} > 0.4\,.
\label{eq:cuts}
\end{eqnarray}
In these expressions, $R$ is the usual boost-invariant angular distance,
$R_{ab} = [\Delta\phi_{ab}^2+\Delta\eta_{ab}^2]^{1/2}$.
We define jets using the anti-$k_T$ algorithm~\cite{antikT} with
parameter $R = 0.4$.  The jets are ordered in $\pT$,
and are labeled $i,j=1,2,\ldots$ in
order of decreasing transverse momentum $p_{\rm T}$, with jet $1$ being the
leading (hardest) jet. 

In addition, we also consider further cuts, which select the kinematic region
for VBF production of the Higgs-like boson, with the boson decaying into two photons.
We will call these the VBF cuts,
\begin{equation}
 M_{jj} > 400  \hbox{ GeV}\,, \hskip 1.5 cm 
    |\Delta\,\eta_{jj}| > 2.8 \,,
\label{eq:VBFcuts}
\end{equation}
where $M_{jj}$ is the invariant mass of the 
subsystem made up of the two hardest jets, 
and $\Delta\,\eta_{jj}$ is the difference in pseudorapidity 
between the hardest and the second-hardest jets.
We will show distributions both with and without VBF cuts.

We use $\mu_R = \mu_F = \HTpartonic/2$, where
\begin{equation}
\HTpartonic \,\equiv\, \pT^{\gamma_1} + \pT^{\gamma_2} + \sum_m p_{\rm T}^m \,,
\label{HT}
\end{equation}
for the central renormalization and factorization scale in our calculation.
The sum runs over all final-state partons $m$, whether or not they are inside jets that pass the
cuts.  This means that modifications to the cuts will not affect the value
of the matrix element at a given point in phase space.  This 
avoids unwanted dependence on experimental cuts.   We quote scale variation bands
corresponding to varying the scales simultaneously up and down by a factor of two, taking the
maximum and minimum of differential cross sections at the five scales 
$\HTpartonic/2\times (1/2,1/\sqrt{2},1, \sqrt{2},2)$.

The calculation proceeds in two phases: generation of \ntuples, and analysis.  In the first phase,
we generate two sets of \root{}~\cite{ROOT} format \ntuples{} using a looser set of cuts,
\begin{eqnarray}
&& \pT^{\gamma_1} > 25 \hbox{ GeV} \,, \hskip 1.5 cm
\pT^{\gamma_2} > 25 \hbox{ GeV} \,, \hskip 1.5 cm 
|\eta^{\gamma}| < 4.5\,, \hskip 1.5 cm \nn\\
&& \pT^{\jet_1} > 25 \hbox{ GeV}\,, \hskip 1.5 cm 
   \pT^{\jet_2} > 25 \hbox{ GeV}\,. \hskip 1.5 cm 
\label{eq:cuts-generation}
\end{eqnarray}
with a second set also imposing VBF cuts looser than those of \eqn{eq:VBFcuts},
\begin{equation}
M_{jj} > 300  \hbox{ GeV}\,, \hskip 1.5 cm 
    |\Delta\,\eta_{jj}| > 2.0 \,. 
\label{eq:VBFcuts-generation}
\end{equation}
A  second set of \ntuples{} is essential to obtaining reasonable statistical uncertainties for the
latter cuts.

We also study the effect of an additional set of cuts, suggested by the ATLAS collaboration,
which select a window on the diphoton invariant mass centered around 
the Higgs-like boson mass,
\begin{eqnarray}
&& \pT^{\gamma_1} > 0.35\,m_{\gamma\gamma}  \,, \hskip 1.2 cm
\pT^{\gamma_2} > 0.25\,m_{\gamma\gamma} \,, \hskip 1.2 cm 
|y^{\gamma}| < 2.37\,, \hskip 1.2 cm \nn\\
&& \pT^{\jet} > 30 \hbox{ GeV}\,, \hskip 1.2 cm  
R_{\gamma,\jet} > 0.4\,,\hskip 1.2cm
|y^{\jet}| < 4.4\,, \hskip 1.2 cm \nn
  122 \leq m_{\gamma\gamma}\leq 130\,.
\label{eq:ATLAScuts}
\end{eqnarray}
The additional VBF cuts here are the same as those in \eqn{eq:VBFcuts-generation}.

As mentioned above, the \ntuples{} store intermediate results
such as parton momenta and coefficients associated with the event
weights for the events passing the looser cuts.
In the second, analysis, phase of our calculation we impose the cuts of \eqn{eq:cuts}, and
in addition those of \eqn{eq:VBFcuts} on the second set.  
(Alternatively, we impose the cuts of \eqn{eq:ATLAScuts}, and in addition those
of \eqn{eq:VBFcuts-generation} in a parallel set.)
We compute
the total cross section as well as various distributions.  The \ntuples{} 
can be used to study the effects of varying the parton distributions, scale
choices, and experimental cuts.

From an experimental point of view, photons must be isolated from
hadronic radiation.  From a theoretical point of view, we cannot
isolate them completely --- for example, by excluding all radiation
from a cone surrounding the photon axis --- because this would disturb
the cancellation of infrared singularities between virtual and
real-emission contributions.  The Frixione isolation
criterion~\cite{FrixioneCone} offers a compromise between the two
requirements that, unlike other isolation criteria, requires no
additional non-perturbative input in the form of fragmentation
functions.  The Frixione isolation cone depends on the radius
$\delta=[(\phi-\phi_\gamma)^2+(\eta-\eta_\gamma)^2]^{1/2}$ of a cone
surrounding a photon, and fixes a distance-dependent limit $E(\delta)$
on the transverse hadronic energy allowed inside,
 \begin{equation}
   \sum_{p} E_{{\rm T}p}
    \theta(\delta-R_{p\gamma})\leq
     E(\delta)
     {\rm \;\;\;\;\;\;\;\;\;\;\;\;with\;\;\;\;\;\;\;\;\;\;\;\;}
    E(\delta) = E_{\rm T}^{\gamma}\,\epsilon
     \Bigg(
      \frac{1-\cos\delta}{1-\cos\delta_0}
     \Bigg)^n\,,
\end{equation}
with the sum taken over all partons.
In our calculation, we used
$\epsilon=0.5\,$, $\,\delta_0=0.4\,$ and $\,n=1\,$ for the Frixione isolation parameters. 
While the Frixione isolation does not match the experimental photon isolation, the large subtractions
performed in LHC analyses to eliminate pile-up effects mean that the traditional cone isolation does not
match experimental practice very well either.

In our study, we use the MSTW2008 LO and NLO
PDFs~\cite{MSTW2008} at the respective orders.  We use the
zero-momentum-squared value, $\alpha_{\rm EM}(0) = 1/137.036$, for the
electromagnetic coupling.  We use the five-flavor
running $\alpha_s(\mu)$ and the value of $\alpha_s(M_Z)$ supplied with
the parton distribution functions.  We do not apply corrections
due to non-perturbative effects such as those induced by the
underlying event or hadronization.  For comparisons to experimental data it
is important to incorporate these effects. 

\newcommand\Tstrut{\rule{0pt}{2.4ex}}       % top strut
\newcommand\Bstrut{\rule[-1.1ex]{0pt}{0pt}} % bottom strut

%%%%%%%%%%%%%%%%%%%%%%%% TABLE %%%%%%%%%%%%%%%%%%%%%%
\begin{table}[t]
\vskip .4 cm
\centering
\begin{tabular}{||c||c|c|c||}
\hline
Cuts &  LO  & NLO & $gg\rightarrow \gamma\gamma gg$ \\  \hline
\Tstrut\Bstrut Basic &~$2.678(0.003)^{+0.808}_{-0.577}$~&~$3.23(0.03)^{+0.31}_{-0.36}$~&~$0.0509(0.0007)$~\\ \hline
\Tstrut\Bstrut VBF &~$0.1398(0.0003)^{+0.0541}_{-0.0359}$~&~$0.159(0.002)^{+0.016}_{-0.021}$~&~$0.004(0.001)$~\\ \hline
\Tstrut\Bstrut ATLAS &~$0.0886(0.0005)^{+0.0264}_{-0.0189}$~&~$0.099(0.002)^{+0.007}_{-0.010}$~&~$0.00157(0.00003)$~\\ \hline
\Tstrut\Bstrut ATLAS VBF &~$0.00392(0.00004)^{+0.00153}_{-0.00101}$~&~$0.0046(0.0001)^{+0.0006}_{-0.0006}$~&~$8.9(0.4)\cdot 10^{-5}$~\\ \hline
\end{tabular}
\caption{Total cross sections in picobarns for \YYjj-jet production with various sets of cuts:
basic (\protect\eqn{eq:cuts}), VBF (\protect\eqns{eq:cuts}{eq:VBFcuts}), 
ATLAS (\protect\eqn{eq:ATLAScuts}), and ATLAS VBF (\protect\eqns{eq:ATLAScuts}{eq:VBFcuts-generation}).
The numerical integration uncertainty is given in parentheses, and the scale dependence
is quoted in superscripts and subscripts.  The contributions for the $gg\rightarrow \gamma\gamma gg$ subprocess,
shown in the last column, are included in the NLO value.
}
\label{CrossSectionTable}
\end{table}
%%%%%%%%%%%%%%%%%%%%%%%%%%%%%%%%

\section{Results}

Our results for the total cross sections with the basic cuts of \eqn{eq:cuts}; with
VBF cuts of \eqn{eq:VBFcuts} in addition; or with the ATLAS cuts
of \eqn{eq:ATLAScuts}; and with the VBF cuts of \eqn{eq:VBFcuts-generation} in addition,
are shown in table~\ref{CrossSectionTable}.

%%%%%%%%%%%%% FIGURE %%%%%%%%%%%%%%%%%%
\begin{figure}[ht]
\centerline{
\includegraphics[clip,scale=0.27]{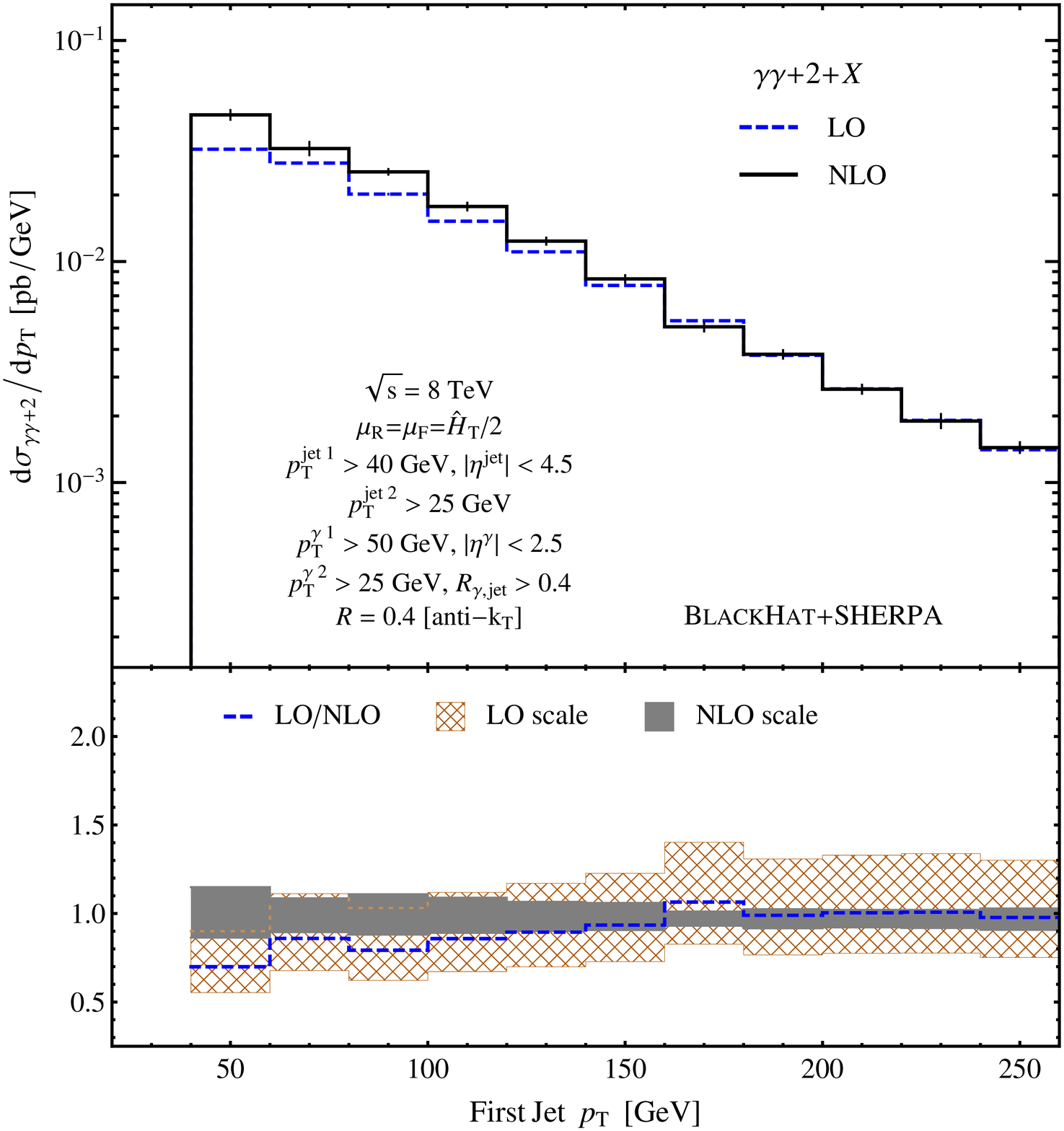}
\includegraphics[clip,scale=0.27]{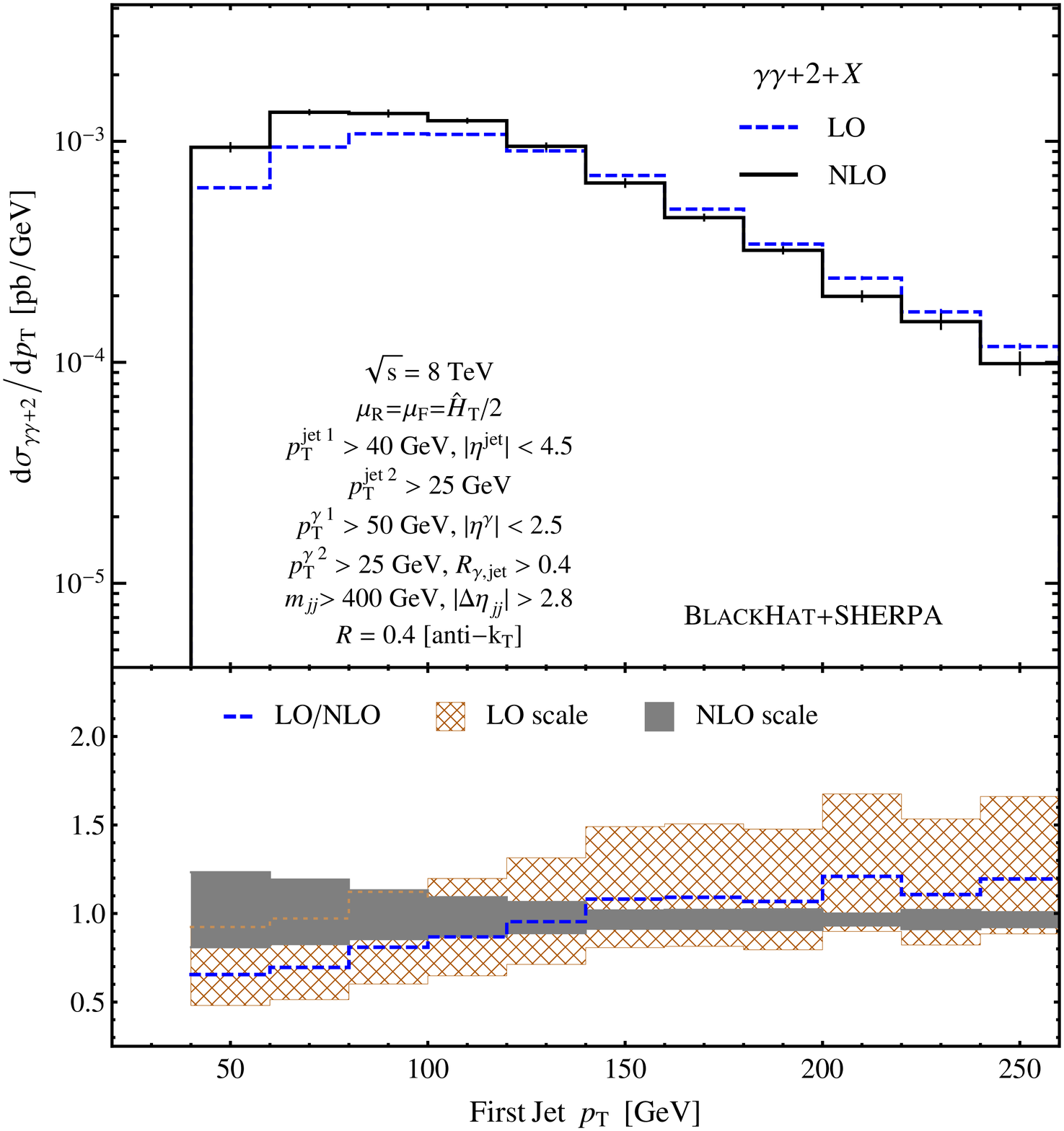}}
\caption{The leading jet's transverse-momentum distribution 
without (left) and with (right) VBF cuts.   The upper panels show the LO (dashed blue) and NLO (solid black) predictions,
while the lower panels show the ratio of these predictions to the NLO prediction, along with the LO (hatched brown)
and NLO (gray) scale uncertainty bands.}
\label{fig:jet1pt}
\end{figure}
%%%%%%%%%%%%%%%%%%%%%%%%%%%%

In Fig.~\ref{fig:jet1pt}, we show the transverse-momentum distribution of the leading jet,
using the cuts in \eqn{eq:cuts}, both before and after the VBF cuts of \eqn{eq:VBFcuts}.
The general features of this distribution before VBF cuts are familiar from studies of
$W$ or $Z$ production in association with jets: the rapidly-falling distribution is slightly
steeper at NLO than at LO, overall the NLO corrections are modest,
 and the scale-dependence bands narrow signficantly at
NLO compared to LO.  

%%%%%%%%%%%%% FIGURE %%%%%%%%%%%%%%%%%%
\begin{figure}[ht]
\centerline{
\includegraphics[clip,scale=0.27]{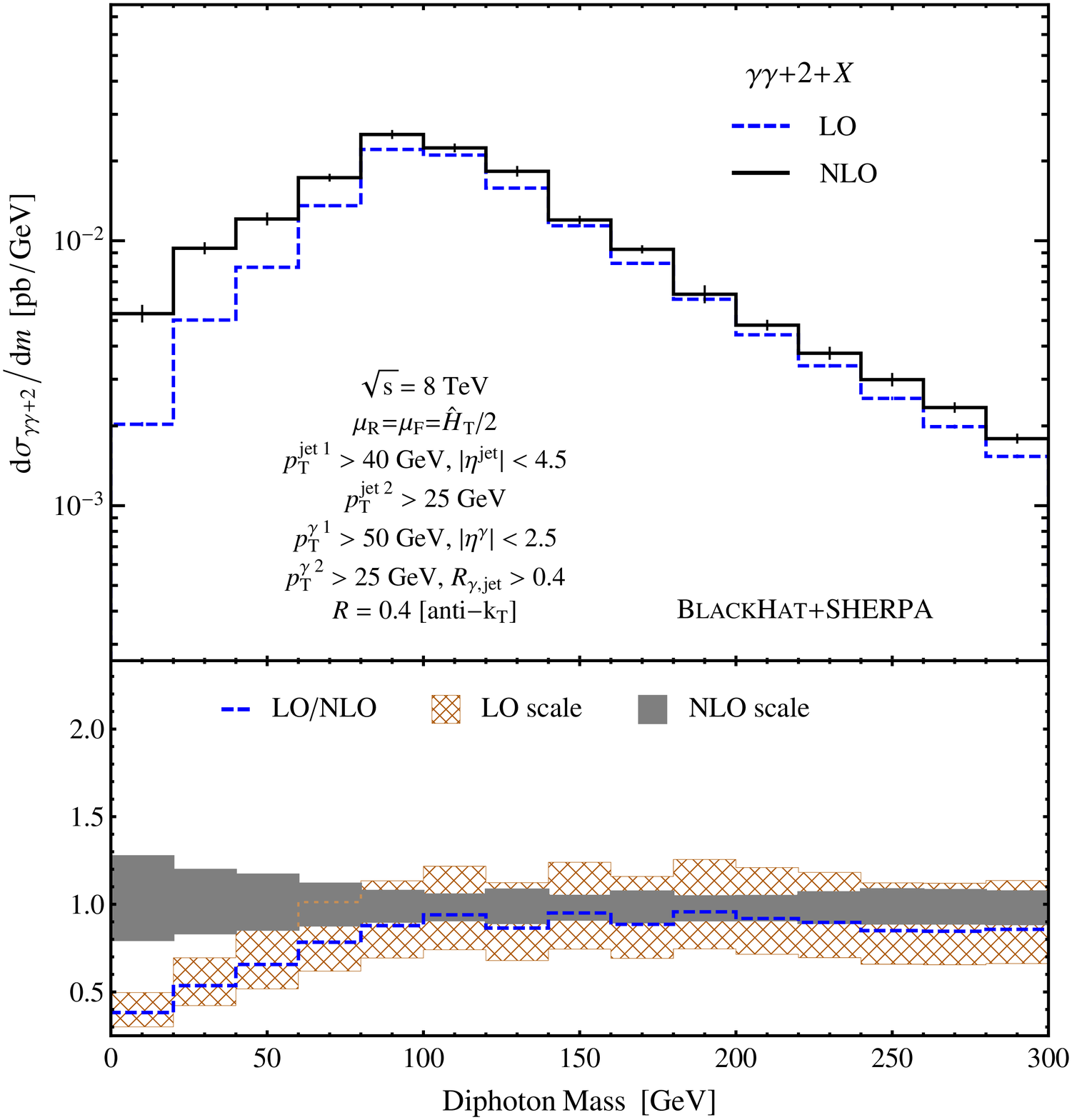}
\includegraphics[clip,scale=0.27]{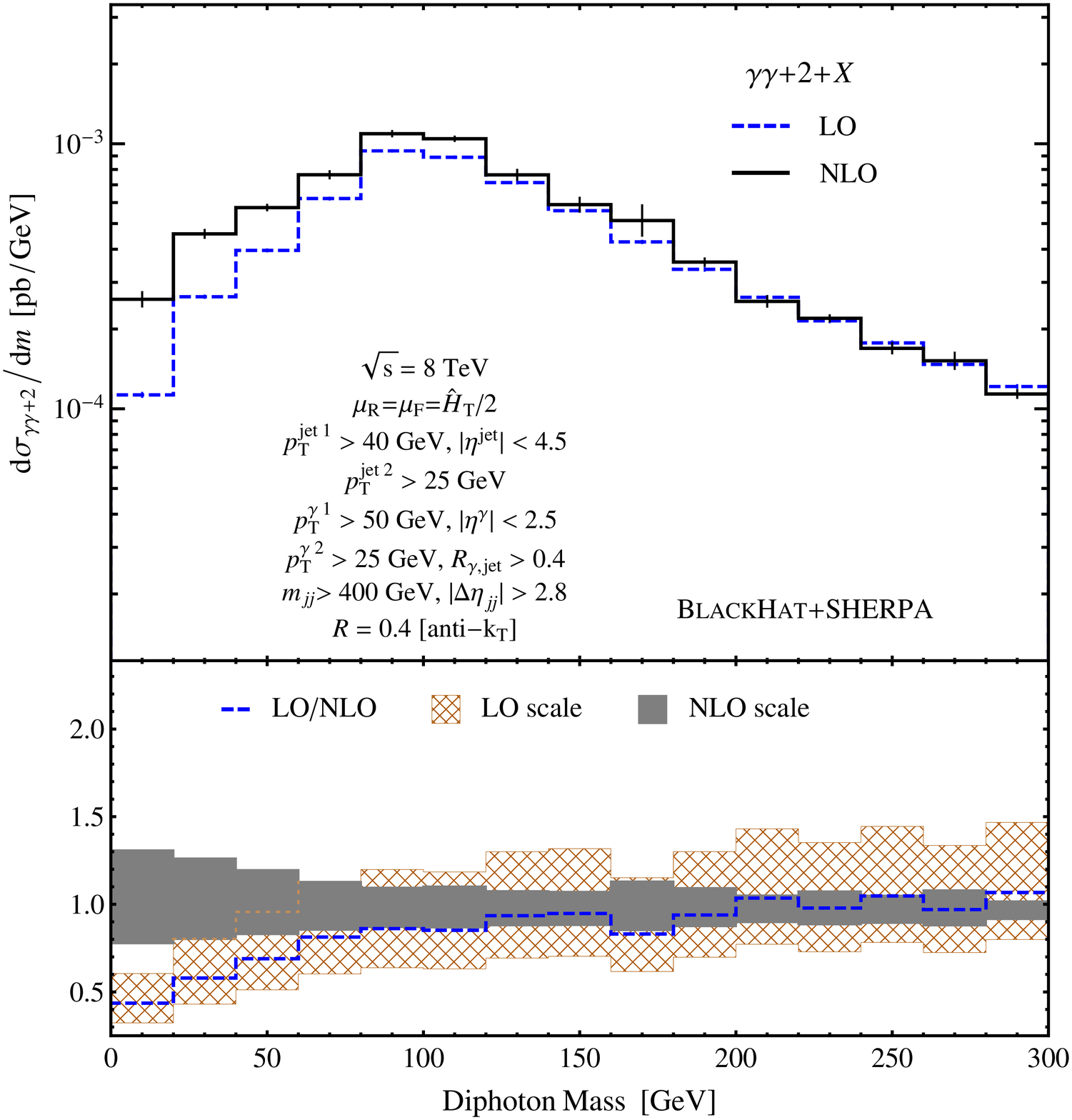}}
\caption{The diphoton invariant mass distribution 
without (left) and with (right) VBF cuts.  The curves and bands are as in Fig.~\protect\ref{fig:jet1pt}.}
\label{fig:diphotonmass}
\end{figure}
%%%%%%%%%%%%%%%%%%%%%%%%%%%%

In Fig.~\ref{fig:diphotonmass}, we show the diphoton invariant mass, using the same cuts.
In these distributions, and also the leading-jet $\pT$ distribution after VBF cuts, we see
that there are sizeable NLO corrections at low values of the observable.  These are presumably
due to relaxation of kinematic constraints with additional radiation; the large corrections are then
accompanied by a widening of the scale-dependence band, as the additional power of 
the strong coupling will cause the real-radiation contribution alone
to have larger scale dependence than the Born contribution.  The NLO scale dependence bands
widen slightly at large diphoton invariant mass, though this effect is eliminated by the VBF cuts.

%%%%%%%%%%%%% FIGURE %%%%%%%%%%%%%%%%%%
\begin{figure}[ht]
\centerline{
\includegraphics[clip,scale=0.27]{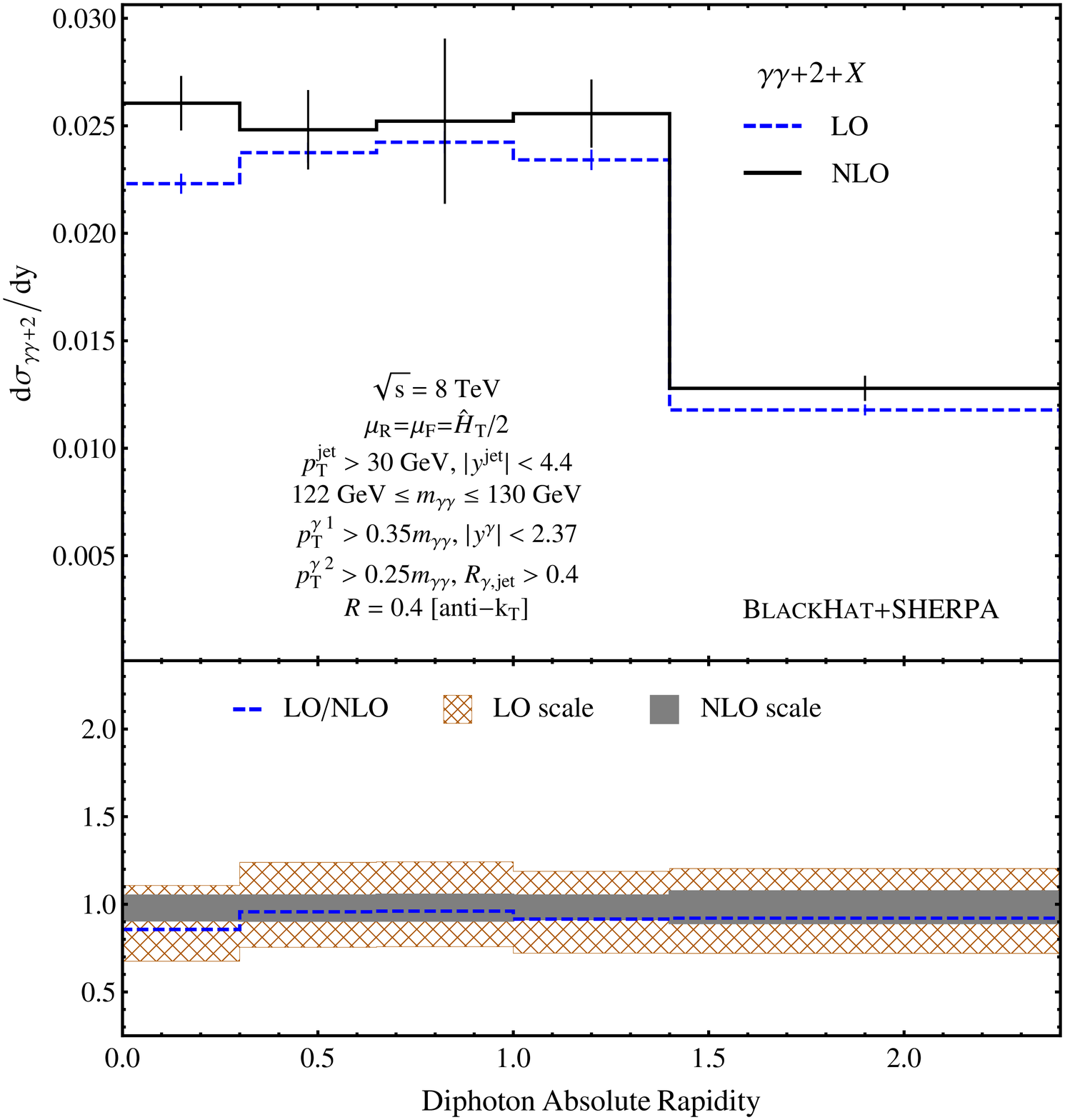}
\includegraphics[clip,scale=0.27]{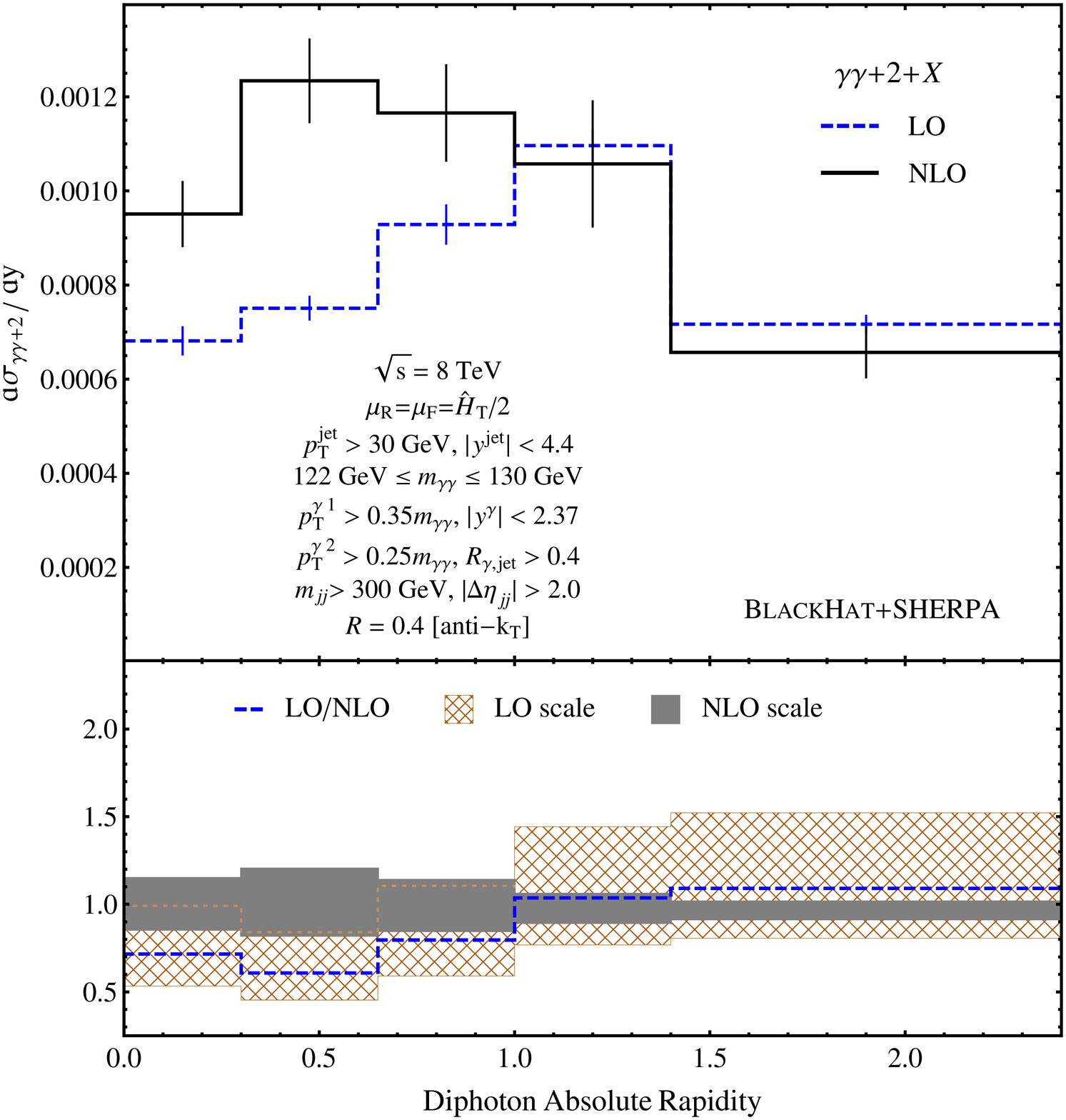}}
\caption{The distribution of the absolute value of the diphoton rapidity, using 
ATLAS-suggested cuts, without (left) and with (right) VBF cuts. The curves and bands are as in Fig.~\protect\ref{fig:jet1pt}.}
\label{fig:ATLASdiphotonrapidity}
\end{figure}
%%%%%%%%%%%%%%%%%%%%%%%%%%%%

%%%%%%%%%%%%% FIGURE %%%%%%%%%%%%%%%%%%
\begin{figure}[ht]
\centerline{
\includegraphics[clip,scale=0.27]{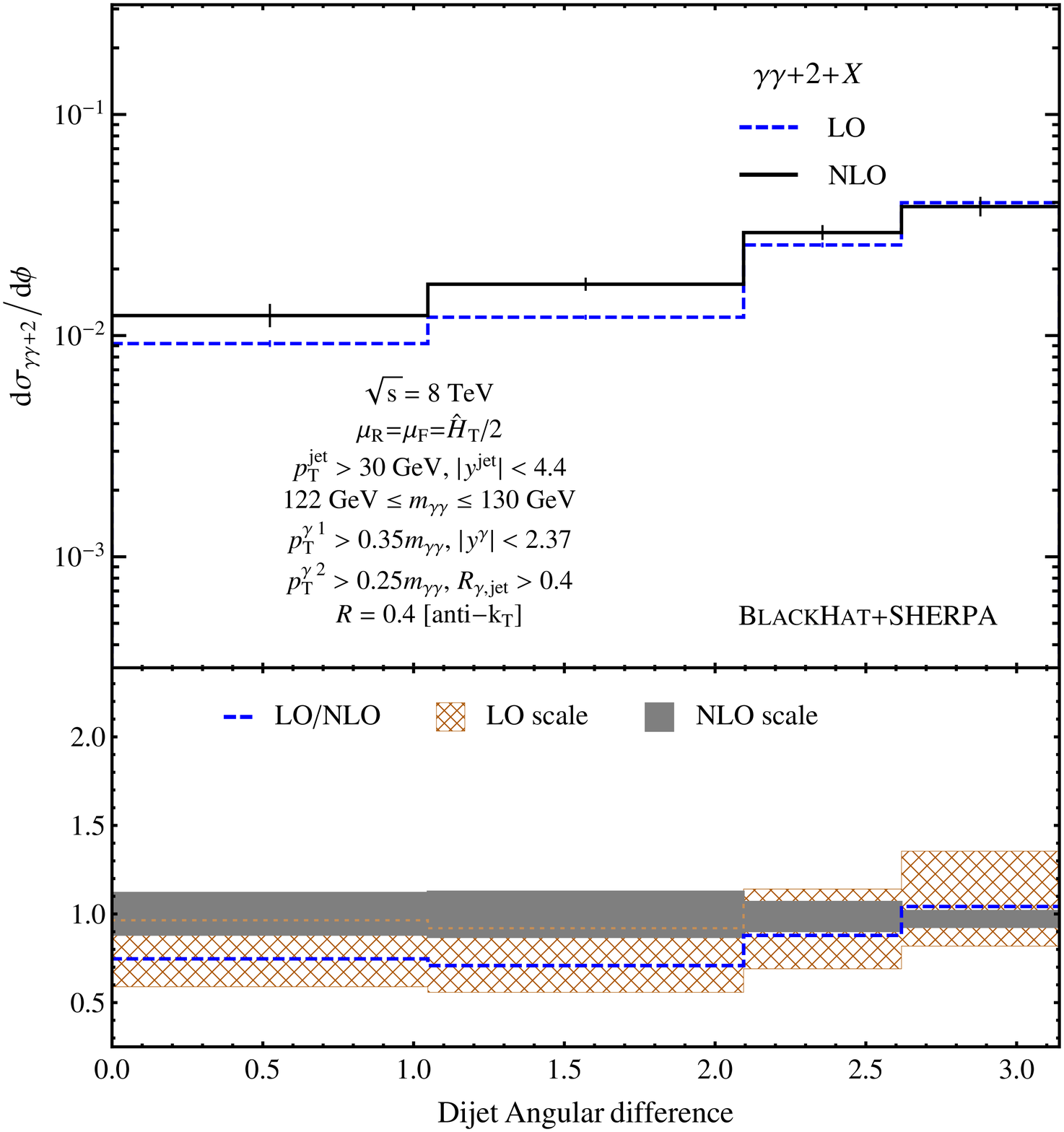}
\includegraphics[clip,scale=0.27]{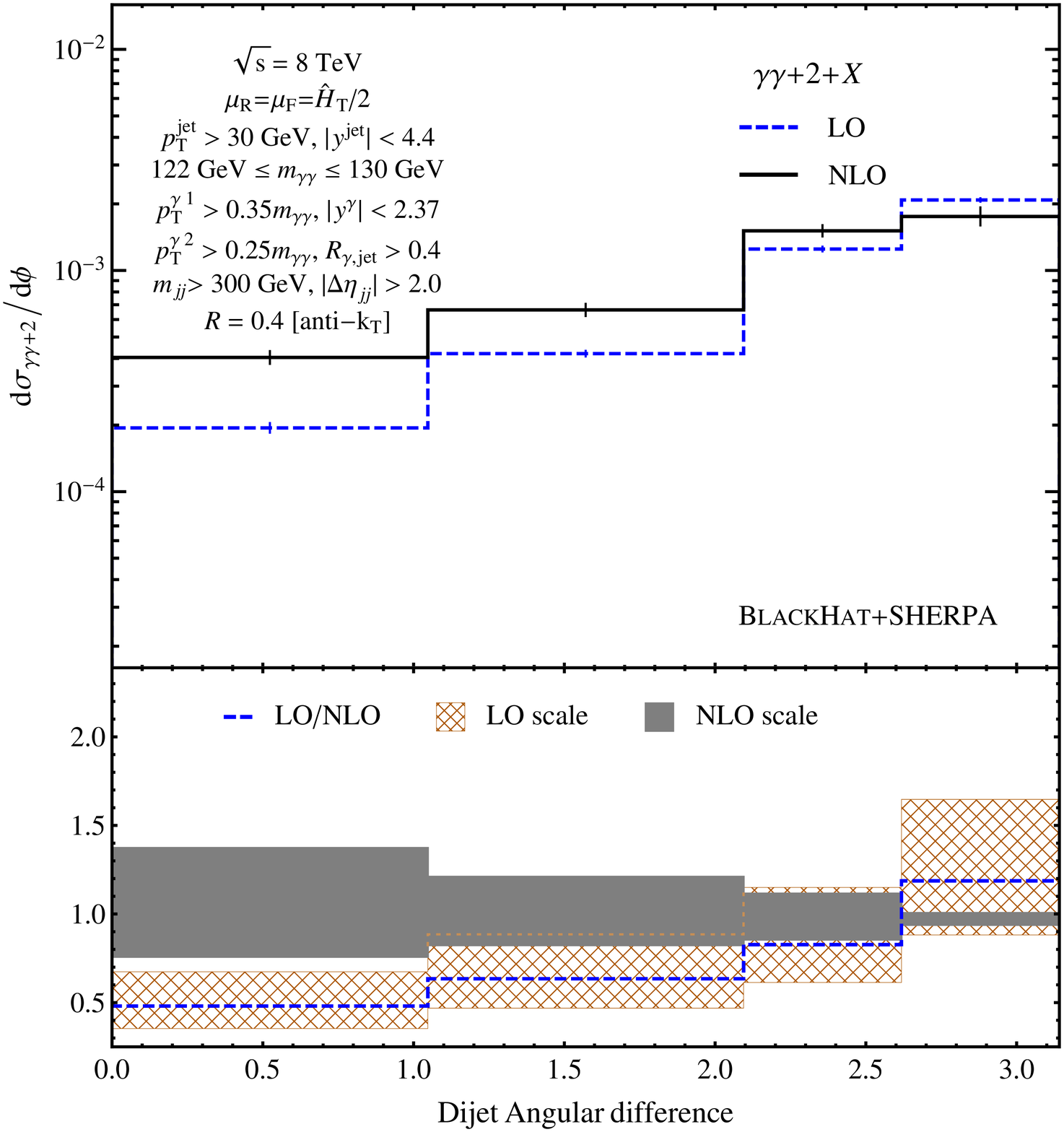}}
\caption{The distribution of the azimuthal angle difference between the two hardest jets, using ATLAS-suggested cuts, 
without (left) and with (right) VBF cuts.  The curves and bands are as in Fig.~\protect\ref{fig:jet1pt}.}
\label{fig:ATLASdijetangle}
\end{figure}
%%%%%%%%%%%%%%%%%%%%%%%%%%%%

In Fig.~\ref{fig:ATLASdiphotonrapidity}, we show the distribution of the absolute value of the diphoton rapidity,
using the cuts in \eqn{eq:ATLAScuts}, both before and after the VBF cuts of \eqn{eq:VBFcuts-generation}.
In Fig.~\ref{fig:ATLASdijetangle}, we show the distribution of the azimuthal angle between the two leading jets,
using the same cuts.  The NLO corrections to both distributions are modest before VBF cuts, with the NLO corrections
favoring emission which allows the two jets to come closer in azimuthal angle.  Here, the VBF cuts do result
in noticeable (to diphoton rapidity) or large (to dijet angular difference) corrections, tending to enhance contributions
with central rapidity for the diphoton pair, and with small azimuthal separation between the leading jets.  
In both regions, the NLO scale dependence is correspondingly larger.

In this contribution, we have presented a computation of \YYjj-jet production at NLO in QCD.  We have given results
for the total cross section and a number of distributions, under a variety of cuts.  The sets of cuts include
those which isolate kinematic regions relevant to vector-boson fusion production of the Higgs-like boson,
to which the process we have studied is an important background.  We leave a number of interesting
issues to future studies.

\section*{Acknowledgments}
We thank Joey Huston for helpful discussions.
This research was supported by the US Department of Energy under
contracts DE--SC0009937 and DE--AC02--76SF00515.  DAK and NALP's research
is supported by the European Research Council under Advanced
Investigator Grant ERC--AdG--228301.  DM's work was supported by the
Research Executive Agency (REA) of the European Union under the Grant
Agreement number PITN--GA--2010--264564 (LHCPhenoNet). 
This research used
resources of Academic Technology Services at UCLA.

%%%%%%%%%%%%%%%%%%%%%

\end{document}